\documentstyle[prb,aps]{revtex}
\input epsfig.sty
\begin{document}

\advance\textheight by 0.2in
\twocolumn[\hsize\textwidth\columnwidth\hsize\csname@twocolumnfalse%
\endcsname

\draft
\begin{flushright}
{\tt to appear in Phys. Rev. B } 
\end{flushright} 
\title{
Current-voltage scaling of chiral and gauge glass models of two-dimensional
superconductors. }
\author{Enzo Granato}
\address{Laborat\'orio Associado de Sensores e Materiais, \\
Instituto Nacional de Pesquisas Espaciais,\\
12201 S\~{a}o Jos\'{e} dos Campos, SP, Brazil}
\maketitle
\draft

\begin{abstract}
The scaling behavior of the current-voltage characteristics of chiral and
gauge glass models of disordered superconductors, are studied numerically,
in two dimensions. For both models , the linear resistance is nonzero at
finite temperatures and the scaling analysis of the nonlinear resistivity is
consistent with a phase transition at $T=0$ temperature characterized by a
diverging correlation length $\xi \propto T^{-\nu _{T}}$ and thermal
critical exponent $\nu _{T}$. The values of $\nu _{T}$ , however, are found
to be different for the chiral and gauge glass models, suggesting different
universality classes, in contrast to the result obtained recently in three
dimensions.
\end{abstract}


]

Considerable attention has been paid recently to the nature of the glass
phase and the universality class of the possible glass transitions in
ceramic or granular high-$T_{c}$ superconductors. In the presence of an
applied magnetic field, a vortex-glass phase transition with vanishing
linear resistance in the low temperature phase has been predicted in three
dimensions \cite{fisher89,FFH} while, in two dimensions, the linear
resistance is nonzero at finite temperatures $T$ but there is, nevertheless,
a zero-temperature transition \cite{hyman} with a divergent correlation
length $\xi $ $\propto T^{-\nu _{T}}$ and critical exponent $\nu _{T}\sim
2$ which determines the behavior of the current-voltage characteristics. The
absence of a finite-temperature vortex-glass transition and current-voltage
scaling has been verified in some two-dimensional superconducting films \cite
{dekker} with a critical exponent $\nu _{T}$ in agreement with theory. In
the theoretical studies, a gauge glass model has been widely used which is
believed to be in the same universality class as the vortex glass. In zero
external field, however, only the standard superconducting transition takes
place. In contrast, in d-wave superconductors materials containing '' $\pi $
'' junctions \cite{rice} a chiral-glass phase has been predicted even at
zero external field in three dimensions based on numerical studies of the
chiral glass model \cite{kawamura,kawali}, or alternatively, and XY spin
glass \cite{villain}. The chiral-glass order parameter is the chirality,
which represents the direction of local current loops introduced by
frustration effects. As for the gauge glass, in two dimensions, the
chiral-glass transition only occurs at zero temperature \cite
{kawamura,Ray,BY96}. It is well known that the chiral glass model has an
additional reflection symmetry which gives rise to quenched in vortices in
the proposed chiral glass phase \cite{villain}. In spite of this, a recent
study of the current-voltage characteristics of the chiral and gauge glass
models in the vortex representation \cite{wengel} found that they exhibit,
within the numerical accuracy, the same critical exponents in three
dimensions, suggesting a common universality class. This would imply that
resistivity measurements alone are not able to distinguish between chiral
and vortex glass states. However, in two dimensions the problem has not yet
been analyzed in detail although several studies have already shown that the
zero-temperature chiral glass transition is characterized by two different
divergent correlation lengths \cite{kawamura,Ray,benakli}, $\xi _{c}$ and $%
\xi _{s}$ , with different critical exponents, $\nu _{c}$ $\sim 2$ and $\nu
_{s}\sim 1$ , describing chiral glass (or vortex glass order) and phase
glass order, respectively. In particular the value of $\nu _{c}$ for the
chiral glass turns out to agree with the exponent for the gauge glass model 
\cite{fisher91}. The question then arises as to what correlation length is
actually probed by nonlinear resistance measurements since this will
determine the current-voltage scaling and the resulting behavior could be
either consistent with $\nu _{T}=\nu _{c}$, the same as the gauge glass
model, or $\nu _{T}=\nu _{s}$ which could serve to identify the
zero-temperature chiral glass transition.

In this work we present a numerical study of the current-voltage
characteristics of chiral and gauge glass models, in two dimensions, in a
representation in terms of the phases of the local superconducting order
parameter. For both models, we find that the linear resistance is nonzero at
finite temperatures and a scaling analysis of the nonlinear resistance is
consistent with a phase transition at $T=0$ temperature characterized by a
diverging correlation length $\xi \propto T^{-\nu _{T}}$ and thermal
critical exponent $\nu _{T}$, in agreement with previous work. The values of 
$\nu _{T}$ , however, are found to be different for the chiral and gauge
glass models, suggesting that measurements of nonlinear resistance probe
mainly the phase correlation length and the models are in different
universality classes, in contrast to the result obtained in three dimensions
in the vortex representation \cite{wengel}. Thus, in two dimensions,
measurements of nonlinear resistance could, in principle, be used to
identify a possible chiral glass in two-dimensional ceramic superconductors.

The chiral glass and gauge glass models can be described by the same
Hamiltonian

\begin{equation}
H=-J_{o}\sum_{<ij>}\cos (\theta _{i}-\theta _{j}-A_{ij})  \label{hamilt}
\end{equation}
where $\theta _{i}$ is the phase of the superconducting order parameter of a
''grain ''at site $i$ of a regular lattice, $J_{o}>0$ is a constant
Josephson coupling and screening effects have been ignored . In the
gauge-glass model \cite{fisher91,wengel}, $\ A_{ij}$ represents a quenched
line-integral of the vector potential which is taken to be uniformly
distributed in the interval $[0,2\pi ]$ , representing the combined effect
of disorder and the external magnetic field , while in the chiral-glass
model \cite{kawamura} $A_{ij}$ has a binary distribution, $0$ or $\pi $ ,
with equal probability, which may represent the phase shift across Josephson
junctions in models of d-wave ceramic superconductors even the absence of
magnetic field\cite{dominguez}. Alternatively, the chiral glass model is
just another representation of the XY spin glass \cite
{villain,kawali,Ray,BY96} with random couplings $J_{ij}=$ $\pm \ J_{o}$ .

To study the current-voltage characteristics of disordered superconductors
described by the Hamiltonian of Eq. (\ref{hamilt}), we assume a resistively
shunted Josephson-junction (RSJ) model for the current flow between grains 
\cite{Shenoy} and use an overdamped Langevin dynamics \cite{falo} to
simulate the nonequilibrium behavior. The Langevin equations can be written
as

\begin{eqnarray}
C_{o}\frac{d^{2}\theta _{i}}{dt^{2}}+\frac{1}{R_{o}}\sum_{j}\frac{d(\theta
_{i}-\theta _{j})}{dt}&=&-J_{o}\sum_{j}\sin (\theta _{i}-\theta
_{j}-A_{ij}) \nonumber \\
&& +I_{i}^{ext}+\sum_{j}\eta _{ij}\ ,  \label{lang}
\end{eqnarray}
where $I^{ext}$ is the external current, $\eta _{ij}$ represents Gaussian
thermal fluctuations satisfying

\begin{eqnarray}
&<&\eta _{ij}(t)>=0 \\
&<&\eta _{ij}(t)\ \eta _{kl}(t^{\prime })>=\frac{2k_{B}T}{R_{o}}\delta
_{ij,kl}\delta (t-t^{\prime })
\end{eqnarray}
and a capacitance to the ground $C_{o}$ is allowed, in addition to the shunt
resistance $R_{o}$ , in order to facilitate the numerical integration \cite
{falo}. We use units where $
\rlap{\protect\rule[1.1ex]{.325em}{.1ex}}h%
/2e=1$ , $R_{o}=1$ , $J_{o}=1$ and set the parameter $JR_{o}^{2}C_{o}=0.5$
in the simulations, corresponding to the overdamped regime. The above
equations were integrated numerically using, typically, a time step $\delta
t=0.02-0.05$ $\tau $ ($\tau =1/R_{o}J_{o}$), time averages computed with $%
2-4\times 10^{5}$ time steps and the results averaged over $5-10$ different
realizations of the disorder. To determine the nonlinear resistivity (or
resistance in two dimensions), $\rho _{nl}=E/J$ , an external current $I$ is
injected uniformly with density $J=I/L$ along one edge of a square lattice
of size $L\times L$ and extracted at the opposite one, with periodic
boundary conditions in the transverse direction. The average voltage drop $V$
across the system is computed as

\begin{equation}
V=\frac{1}{L}\sum_{j=1}^{L}(\frac{d\theta _{1,j}}{dt}-\frac{d\theta _{L,j}}{%
dt})
\end{equation}
and the average electric field by $E=V/L$ . We have also computed the linear
resistance, $R_{L}=\lim_{J\rightarrow 0}E/J$ , without finite current
effects, directly from the long-time equilibrium fluctuations of the phase
difference across the system $\Delta \theta (t)=\sum_{j=1}^{L}(\theta
_{1,j}-\theta _{L,j})/L$ as

\begin{equation}
R_{L}=\frac{1}{2T}(\Delta \theta (t))^{2}/t  \label{linear}
\end{equation}
which can be obtained from Kubo formula of equilibrium voltage-voltage
fluctuations, $R_{L}=\frac{1}{2T}\int dt<V(t)V(0)>$ , using the Josephson
relation $V=d\theta /dt$ . Lattices of sizes $L=$ $16$ , $24$ and $34$ were
used in the simulations with the main results obtained for the largest
system size.

\begin{figure}[tbp]
\centering\epsfig{file=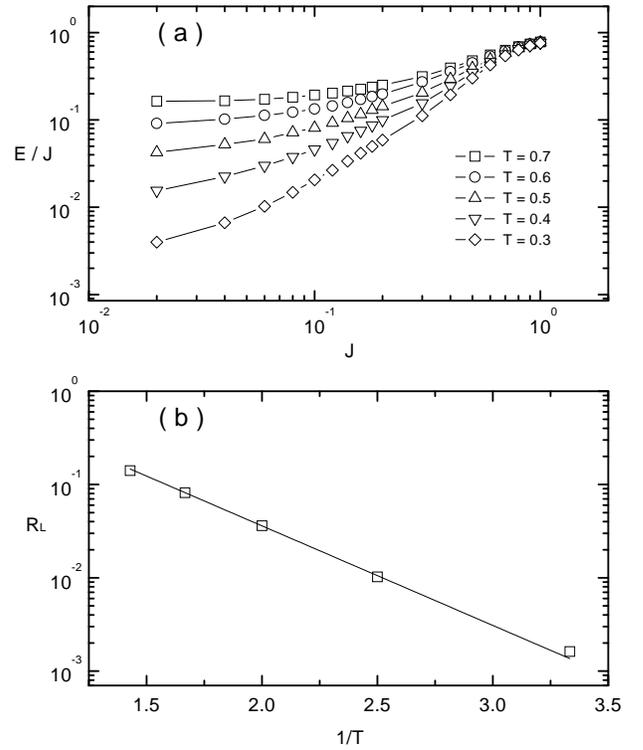,bbllx=1cm,bblly=1cm,bburx=20cm,
 bbury=28cm,width=8.5cm}
\caption{(a) Nonlinear resistance $E/J$ as a function of temperature $T$ for
the gauge glass model. (b)Arrhenius plot for the temperature dependence of
the linear resistance $R_L$. }
\label{resgg}
\end{figure}

\begin{figure}[tbp]
\centering\epsfig{file=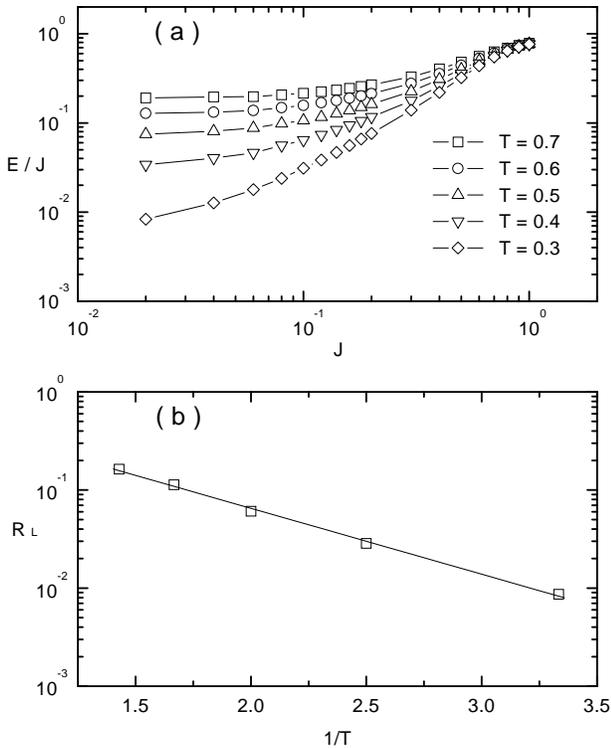,bbllx=1cm,bblly=1cm,bburx=20cm,
 bbury=28cm,width=8.5cm}
\caption{ (a) Nonlinear resistance $E/J$ as a function of temperature $T$
for the chiral glass model. (b)Arrhenius plot for the temperature dependence
of the linear resistance $R_L$. }
\label{rescg}
\end{figure}

The nonlinear resistance $E/J$ as a function of current density $J$ and
temperature $T$ for the gauge glass is shown in Fig. \ref{resgg}a and an
Arrhenius plot of the linear resistance $R_{L}$ in Fig.\ref{resgg}b. They
are consistent with previous results obtained in smaller systems \cite{hyman}%
. The corresponding results for the chiral glass are shown Figs. \ref{rescg}%
a and \ref{rescg}b . The data for both models show the expected behavior for
a $T=0$ superconducting transition \cite{FFH,hyman}. In Figs. \ref{resgg}a
and \ref{rescg}a, the ratio $E/J$ tends to a finite value for small $J$,
corresponding to the linear resistance $R_{L},$ which depends strongly on
the temperature. This is confirmed in Figs. \ref{resgg}b and \ref{rescg}b
where the linear resistance obtained at $J=0$ from Eq. (\ref{linear}) is
consistent with an exponential decrease with temperature indicating a finite
energy barrier for vortex motion. For increasing $J$ , there is a smooth
crossover to nonlinear behavior that appears at smaller currents for
decreasing temperatures. If one assumes a $T=0$ transition with a power-law
divergent correlation length $\xi _{T}\propto T^{-\nu _{T}}$ and since
the external current density $J$ introduces and additional length scale $%
l\sim kT/J,$ the behavior of the nonlinear resistivity normalized to $R_{L}$
can be cast into the scaling form \cite{FFH,hyman}

\begin{equation}
\frac{E}{JR_{L}}=g(\frac{J}{T^{1+\nu _{T}}})  \label{scaling}
\end{equation}
where $g$ is a scaling function and $g(0)=1,$ which contains a single
parameter, the critical thermal exponent $\nu _{T}$ . This scaling form
implies that the characteristics current density $J_{nl}$ at which nonlinear
behavior is expected to set in decreases with temperature as a power law $%
J_{nl}\propto T^{1+\nu _{T}}$ .

\begin{figure}[tbp]
\centering\epsfig{file=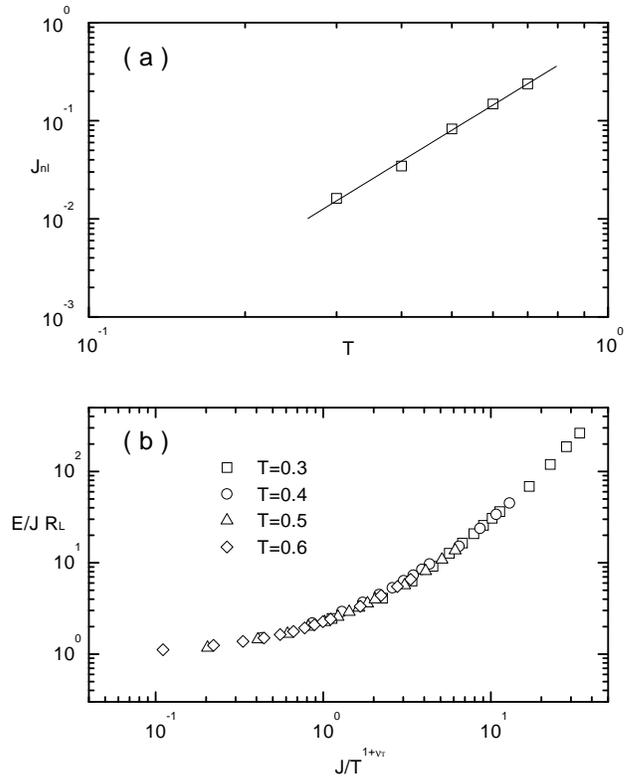,bbllx=1cm,bblly=1cm,bburx=20cm,
 bbury=28cm,width=8.5cm}
\caption{(a) Crossover current density $J_{nl}$ as a function of temperature
for the gauge glass model. (b) Scaling plot $E/J R_L$ $\times$ $%
J/T^{1+\nu_T} $ for $\nu_T=2.35$.}
\label{scalgg}
\end{figure}

\begin{figure}[tbp]
\centering\epsfig{file=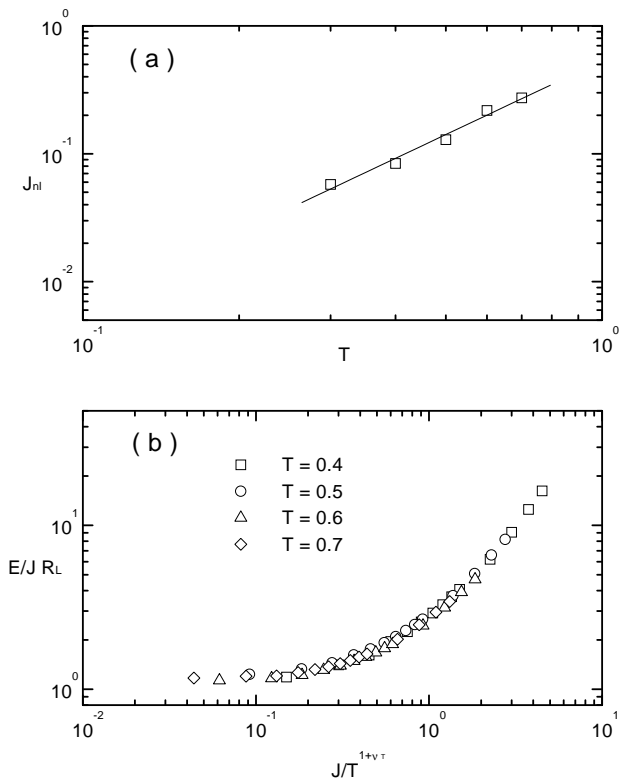,bbllx=1cm,bblly=1cm,bburx=20cm,
 bbury=28cm,width=8.5cm}
\caption{(a) Crossover current density $J_{nl}$ as a function of temperature
for the chiral glass model. (b) Scaling plot $E/J R_L$ $\times$ $%
J/T^{1+\nu_T}$ for $\nu_T=1.2$.}
\label{scalcg}
\end{figure}

We now proceed to verify the scaling hypothesis and obtain a numerical
estimate of the critical exponent $\nu _{T}$ for the gauge and chiral glass
models. Fig. \ref{scalgg}a shows the temperature dependence of $J_{nl}$ for
the gauge glass, defined as the value of $J$ where $E/JR_{L}$ starts to
deviate from a fixed value, chosen to be $2$ . The behavior is consistent
with a power-law behavior and the slope of the loglog plot provides a direct
estimate of $\nu _{T}=2.2(2)$ . A scaling plot according to Eq. (\ref
{scaling}) for the gauge glass is shown Fig. \ref{scalgg}b obtained by
adjusting the parameter $\nu _{T}$ so that a best data collapse is obtained.
The data collapse supports the scaling behavior of Eq. (\ref{scaling}) and
provides an independent estimate of $\nu _{T}=2.4$ . From the two
independent estimates we finally get $\nu _{T}^{gg}=2.3(2)$, a result
consistent with a similar analysis of the nonlinear resistivity obtained
from the coulomb-gas representation of the gauge glass model \cite{hyman}.
Following the same procedure for the chiral glass model, we obtain $\nu
_{T}=0.9(2)$ from the loglog plot in Fig. \ref{scalcg}a and $\nu _{T}=1.2$
for the best data collapse in Fig. \ref{scalcg}b, giving a final estimate $%
\nu _{T}^{cg}=1.1(2)$ .

The values of $\nu _{T}$ for the gauge glass and chiral glass models
obtained from the above analysis are quite different from each other
suggesting different universality classes for the $T=0$ transition. This is
consistent with the additional reflection symmetry property of the chiral
glass \cite{villain,kawamura}, where changing $\theta _{i}\rightarrow
-\theta _{i}$ leaves the Hamiltonian of Eq. (\ref{hamilt}) unchanged,
whereas for the gauge glass there is only a continuous symmetry. In fact,
other studies of the chiral glass \cite{kawamura,Ray,BY96} find that the $%
T=0 $ transition is characterized by two different correlation lengths with
different critical exponents, $\xi _{c}\propto T^{-\nu _{c}}$ describing
chiral glass order and $\xi _{s}\propto T^{-\nu _{s}}$ describing phase
glass (or alternatively, XY spin glass) order with the estimates $\nu
_{c}\sim 2.0$ and $\nu _{s}\sim 1.0$ . Since the relevant length scale for
phase slippage, which leads to voltage fluctuations, at finite temperatures
is given by $\xi _{s}$ we expect that the nonlinear resistivity scaling can
be described by Eq. (\ref{scaling}) with $\nu _{T}=\nu _{s}$ which agrees
with our numerical result for $\nu _{T}^{cg}$ . A more complicated scaling
analysis is also possible involving two correlation lengths but since $\xi
_{s}<<\xi _{c}$ at low temperatures, $\xi _{s}$ should dominate the
resistive behavior and a scaling analysis with a single length scale is a
reasonable approximation.

The distinct critical exponents found for the chiral and gauge glass models
in two dimensions is in sharp contrast with the result obtained for the same
models in three dimensions by Wengel and Young \cite{wengel} where the
critical exponents at the finite temperature transition agree within errors
suggesting a common universality class. This implies that in three
dimensions, resistivity measurement alone cannot distinguish between chiral
and gauge glass states. However, our results suggest that, at least in two
dimensions, measurements of nonlinear resistivity probe mainly the phase
correlation length which has different critical exponents for the chiral and
gauge glass and could, in principle, be used to identify, as has been done
for the vortex glass \cite{dekker}, a possible chiral glass in
two-dimensional ceramic superconductors.

\smallskip

This work was supported by FAPESP (Proc. 97/07250-8). Allocation of computer
time at the Plasma Group (LAP/INPE) is gratefully acknowledged.

\end{document}